\documentclass[%
reprint,
showpacs,preprintnumbers,
nofootinbib,
amsmath,amssymb,
aps,
prd,
floatfix,
]{revtex4-1}
\pdfoutput=1
\usepackage{graphicx}
\usepackage{dcolumn}
\usepackage{footnote}
\usepackage{bm}
\usepackage{color}
\newcommand{\nua}[1]{\ensuremath{\rlap{\kern-2.5pt\ensuremath{\overset{\scriptscriptstyle(-)}{\phantom{\nu}}}}{\ensuremath{{\nu}_{#1}}}}}
\begin{document}

\title{Light sterile neutrino sensitivity of $^{163}$Ho experiments}

\author{L. Gastaldo$^{1}$}
\author{C. Giunti$^{2}$}
\author{E. M. Zavanin$^{2,3,4}$}
\affiliation{
$^1$
Kirchhoff Institute for Physics, Heidelberg University, INF 227, 69120 Heidelberg, Germany
\\
$^2$
INFN, Sezione di Torino, Via P. Giuria 1, I--10125 Torino, Italy
\\
$^3$
Department of Physics, University of Torino, Via P. Giuria 1, I--10125 Torino, Italy
\\
$^4$
Instituto de F\'\i sica Gleb Wataghin, Universidade Estadual de Campinas - UNICAMP,
\\
Rua S\'ergio Buarque de Holanda, 777, 13083-859 Campinas SP Brazil
}

\date{18 May 2016}

\begin{abstract}
We explore the sensitivity of $^{163}$Ho electron capture experiments
to neutrino masses in the standard framework of three-neutrino mixing
and in the framework of 3+1 neutrino mixing with a sterile neutrino
which mixes with the three standard active neutrinos,
as indicated by the anomalies found in
short-baseline neutrino oscillations experiments.
We calculate the sensitivity to neutrino masses and mixing
for different values
of the energy resolution of the detectors,
of the unresolved pileup fraction
and
of the total statistics of events,
considering the expected values of these parameters in the two
planned stages of the ECHo project
(ECHo-1k and ECHo-1M).
We show that an extension of the ECHo-1M experiment
with the possibility to collect $10^{16}$ events
will be competitive with the KATRIN experiment.
This statistics will allow to explore part of the 3+1
mixing parameter space indicated by the global analysis of
short-baseline neutrino oscillation experiments.
In order to cover all the allowed region,
a statistics of about $10^{17}$ events will be needed.
\end{abstract}

\maketitle

\section{Introduction}
\label{sec:intro}

The observation of neutrino oscillations is a clear demonstration that neutrinos are massive particles. The data of solar, atmospheric and long-baseline
neutrino oscillation experiments
are explained in the standard scheme of three-neutrino mixing (3$\nu$)
in which the three active neutrinos
$\nu_{e}$,
$\nu_{\mu}$,
$\nu_{\tau}$
are unitary linear combinations of the three massive neutrinos
$\nu_{1}$,
$\nu_{2}$,
$\nu_{3}$,
with respective masses
$m_{1}$,
$m_{2}$,
$m_{3}$
(see Refs.~\cite{Bellini:2013wra,Wang:2015rma}).
A global analysis of the data of solar, atmospheric and long-baseline
neutrino oscillation experiments
\cite{Forero:2014bxa,Bergstrom:2015rba,Capozzi:2016rtj}
leads to an accurate determination of the three mixing angles
and of the two independent solar and atmospheric squared-mass differences,
$\Delta{m}^2_{\text{SOL}} = \Delta{m}^2_{21} \simeq 7.4 \times 10^{-5} \, \text{eV}^2$
and
$\Delta{m}^2_{\text{ATM}} = |\Delta{m}^2_{31}|
\simeq |\Delta{m}^2_{32}|
\simeq 2.50 \times 10^{-3} \, \text{eV}^2$
\cite{Capozzi:2016rtj},
with
$\Delta{m}^2_{kj} \equiv m_{k}^2 - m_{j}^2$.

The 3$\nu$ paradigm is presently challenged by anomalies
found in short-baseline (SBL) neutrino oscillation experiments:
the reactor antineutrino anomaly
\cite{Mueller:2011nm,Mention:2011rk,Huber:2011wv},
which is a deficit of the rate of $\bar\nu_{e}$ events measured
in reactor neutrino experiments;
the Gallium neutrino anomaly
\cite{Abdurashitov:2005tb,Laveder:2007zz,Giunti:2006bj,Giunti:2010zu,Giunti:2012tn},
consisting in a deficit of the rate of $\nu_{e}$ events measured
in the Gallium radioactive source experiments
GALLEX
\cite{Kaether:2010ag}
and
SAGE
\cite{Abdurashitov:2009tn};
the LSND anomaly,
which is an excess of the rate of $\bar\nu_{e}$ events
in a beam composed mainly of $\bar\nu_{\mu}$'s produced
by $\mu^{+}$ decay at rest
\cite{Athanassopoulos:1995iw,Aguilar:2001ty}.
These anomalies
cannot be explained by neutrino oscillations in the 3$\nu$ scenario.
A possible explanation,
still in the framework of neutrino oscillations,
requires
the existence of a new short-baseline squared-mass difference
$\Delta{m}^2_{\text{SBL}} \gtrsim 1 \, \text{eV}^2$,
which is much larger than
the solar and atmospheric squared-mass differences.
The new short-baseline squared-mass difference
requires the existence of at least one new massive neutrino
$\nu_{4}$ with mass $m_{4}$
such that
$\Delta{m}^2_{\text{SBL}} = |\Delta{m}^2_{41}|$
(see the review in Ref.~\cite{Gariazzo:2015rra}).
In the flavor basis there must be a sterile neutrino $\nu_{s}$
and the mixing of the left-handed neutrino fields is given by
\begin{equation}
\nu_{\alpha L}
=
\sum_{k=1}^{4}
U_{\alpha k} \nu_{k L}
\qquad
(\alpha=e,\mu,\tau,s)
,
\label{mixing}
\end{equation}
where $U$ is the unitary $4\times4$ mixing matrix.
In this so-called 3+1 scenario
the new massive neutrino must be mainly sterile in order not to spoil the
fit of the data of solar, atmospheric and long-baseline experiments
(see the reviews in Refs.~\cite{Bilenky:1998dt,GonzalezGarcia:2007ib,Abazajian:2012ys,Conrad:2012qt,Palazzo:2013me,Gariazzo:2015rra}):
\begin{equation}
|U_{\alpha 4}| \ll 1
\quad
\text{for}
\quad
\alpha=e,\mu,\tau
.
\label{perturbation}
\end{equation}
In other words, the 3+1 scheme must be a perturbation
of the standard three-neutrino mixing.

Several experiments are planned to check the existence
of eV sterile neutrinos
(see the reviews in Refs.~\cite{Lasserre:2015eva,Lhuillier:2015fga,Caccianiga:2015ega,Spitz:2015gga,Gariazzo:2015rra,Giunti:2015wnd,Stanco:2016gnl,Fava201652})
with high-precision investigations of neutrino oscillations over short
baselines by using very accurate detectors
for investigating the disappearance of reactor electron antineutrinos
(DANSS \cite{Danilov:2014vra},
NEOS \cite{Kim:2015qlu},
Neutrino-4 \cite{Serebrov:2013yaa},
PROSPECT \cite{Ashenfelter:2015uxt},
SoLid \cite{Ryder:2015sma},
STEREO \cite{Helaine:2016bmc})
and electron neutrinos produced by very intense radioactive sources
(BEST \cite{Barinov:2016znv},
CeSOX \cite{Borexino:2013xxa}).
New accelerator experiments will perform robust
investigations of short-baseline
$\nua{\mu}\to\nua{e}$
transitions
(JSNS2 \cite{Harada:2016vlb},
SBN \cite{Antonello:2015lea})
and
$\nua{\mu}$ disappearance
(KPipe \cite{Axani:2015zxa},
SBN \cite{Antonello:2015lea}).
Moreover,
there is an increasing interest in the study of the
effects of light sterile neutrinos
in neutrinoless double-$\beta$ decay experiments
\cite{Barry:2011wb,Li:2011ss,Rodejohann:2012xd,Giunti:2012tn,Girardi:2013zra,Pascoli:2013fiz,Meroni:2014tba,Abada:2014nwa,Giunti:2015kza,Pas:2015eia},
in solar neutrino experiments
\cite{Palazzo:2011rj,Palazzo:2012yf,Giunti:2012tn,Palazzo:2013me,Long:2013hwa,Long:2013ota,Kopp:2013vaa},
in long-baseline neutrino oscillation experiments
\cite{deGouvea:2014aoa,Klop:2014ima,Berryman:2015nua,Gandhi:2015xza,Palazzo:2015gja,Agarwalla:2016mrc,Agarwalla:2016xxa,Choubey:2016fpi,Agarwalla:2016xlg},
in atmospheric neutrino experiments
\cite{Razzaque:2011ab,Razzaque:2012tp,Gandhi:2011jg,Esmaili:2012nz,Esmaili:2013cja,Esmaili:2013vza,Rajpoot:2013dha,Lindner:2015iaa,Liao:2016reh,TheIceCube:2016oqi}
and in cosmology
(see Refs.~\cite{Lesgourgues-Mangano-Miele-Pastor-2013,RiemerSorensen:2013ih,Archidiacono:2013fha,Lesgourgues:2014zoa,Gariazzo:2015rra,Hannestad:2016mvv}).

Although the data of short-baseline experiments can be explained either with
$m_{1},m_{2},m_{3} < m_{4}$
or
$m_{4} < m_{1},m_{2},m_{3}$,
the second case is strongly disfavored by cosmological measurements
\cite{Ade:2015xua}
and by the experimental bounds on
neutrinoless double-$\beta$ decay
(assuming that massive neutrinos are Majorana particles;
see Ref.~\cite{Bilenky:2014uka}),
which favor a scenario with
$m_{1},m_{2},m_{3} \ll m_{4}$.
In this paper we consider this scenario,
which implies that
$m_{4}^2 \simeq \Delta{m}^2_{41} = \Delta{m}^2_{\text{SBL}} \gtrsim 1 \, \text{eV}$.
This relation allows us to compare the results
of the experiments measuring directly $m_{4}$
with the results of short-baseline neutrino oscillation
experiments.

The fact that a heavy massive neutrino $\nu_{4}$ is mixing with the three light massive neutrinos to compose the electron neutrino can give a very clear fingerprint in the spectra of nuclear beta decay and electron capture.
This means that experiments designed for the direct investigation of the electron (anti-)neutrino mass have the possibility to scrutinize the parameter space of active-sterile neutrino mixing indicated by short-baseline experiments. The evidence for the existence of such a sterile neutrino would be a kink in the spectrum positioned at $Q - m_4$
\cite{Nakagawa:1963uw,Shrock:1980vy,Shrock:1981ct}, where $Q$ is the energy available to the decay, which is given by the difference between the masses of the parent and daughter atoms.
The amplitude of this kink is related to the mixing
$|U_{e4}|$ that $\nu_4$ has with $\nu_{e}$.

Presently there are two nuclides
which are used for the direct investigation of neutrino masses\footnote{
Note that
the $^3$H beta-decay process
is sensitive to the antineutrino masses,
whereas
the $^{163}$Ho electron-capture process
is sensitive to the neutrino masses.
Hence,
the comparison of the experimental results of the two processes
is a test of the CPT symmetry,
which implies the equality
of neutrino and antineutrino masses.
}:
tritium
($^3$H)
undergoing the beta-decay process
${}^3\text{H} \to {}^3\text{He} + e^{-} + \bar\nu_{e}$
and
holmium
($^{163}$Ho)
undergoing the electron-capture process
$e^{-} + {}^{163}\text{Ho} \to {}^{163}\text{Dy} + \nu_{e}$
(see the reviews in Refs.~\cite{Drexlin:2013lha,Dragoun:2015oja,Nucciotti:2015rsl}).
New generation experiments using these nuclides
are expected to reach a sensitivity to sub-eV values of the
effective electron neutrino mass.
Therefore they can investigate the existence of an eV-scale massive neutrino
which has a significant mixing with $\nu_{e}$.
The sensitivity that can be reached by the KATRIN experiment
\cite{Osipowicz:2001sq,Mertens-TAUP2015}
to the signature of $\nu_{4}$ in the $^3$H beta spectrum was studied in
Refs.~\cite{Riis:2010zm,Formaggio:2011jg,SejersenRiis:2011sj,Esmaili:2012vg,Mertens-TAUP2015}.
These works proved that the KATRIN experiment could,
within three years of measuring time and at nominal performance,
rule out a large part of the parameters space required to explain the anomalies in short-baseline experiments.

In this paper we investigate the
sensitivity of $^{163}$Ho electron capture experiments
to neutrino masses in the standard framework of three-neutrino mixing
and in the framework of 3+1 neutrino mixing with an eV-scale sterile neutrino.
We consider in particular
the first two planned phases of the ECHo project,
ECHo-1k and ECHo-1M \cite{Gastaldo:2013wha,Hassel:2016echo}.
Other $^{163}$Ho experimental projects are
HOLMES \cite{Alpert:2014lfa},
which has a program to investigate small neutrino masses
competitive with the ECHo program,
and
NuMECS \cite{Croce:2015kwa},
which at least for the moment is only aiming at
a precise measurement of the
$^{163}$Ho decay spectrum.

The plan of the paper is as follows.
In Section~\ref{sec:Ho} we describe the effect of neutrino masses in
$^{163}$Ho electron capture.
In Section~\ref{sec:ECHo}
we describe the characteristics of the ECHo experiment
which are relevant for our analysis.
In Section~\ref{sec:3nu}
we present our estimation of the sensitivity of the ECHo experiment
to the effective neutrino mass in the $3\nu$ framework.
In Section~\ref{sec:3p1}
we calculate the sensitivity of the ECHo experiment
to $m_{4}$ in the case of 3+1 neutrino mixing
and we compare it with the
region in the space of the mixing parameters
allowed by the global analysis of short-baseline neutrino oscillation data.
In Section~\ref{sec:conclusions} we present our conclusions.

\section{$^{163}$Ho electron capture process}
\label{sec:Ho}

The property that makes $^{163}$Ho the best isotope for investigating the electron neutrino mass is the very small energy $Q$ available to the decay.
Recently, the $Q$-value has been precisely determined by Penning trap mass spectrometry to be
$Q = 2833 \pm 30_{\text{stat}} \pm 15_{\text{syst}}$ eV \cite{Eliseev:2015pda}.
At the present knowledge, this is the lowest $Q$ for all nuclides undergoing electron capture processes.

In an electron capture process one electron from the $^{163}$Ho atomic levels
is captured, leading to a transformation of a proton into a neutron
and the emission of an electron neutrino.
The daughter atom, $^{163}$Dy is left in an excited state which, at the leading order, is described by a hole in the shell from which the electron has been captured and one electron more in the 4$f$ shell with respect to the ones foreseen for the dysprosium atom in the ground state.
The excitation energy can then be released through the emission of x-rays or electrons (Auger or Coster-Kronig transition). We indicate the sum of all the energy released in the electron capture process minus the one taken away by the neutrino as $E_{\text{c}}$. This is the quantity that is measured by calorimetric techniques in modern experiments studying the $^{163}$Ho decay \cite{Gastaldo:2014hga}.
The concept of these experiments was initially proposed more then thirty year ago by De Rujula and Lusignoli \cite{DeRujula:1981ti,DeRujula:1982qt}.

The decay scheme can then be divided in the following two steps:
\begin{align}
&^{163}\text{Ho} \to {}^{163}\text{Dy}^* + \nu_e, \\
&^{163}\text{Dy}^* \to {}^{163}\text{Dy} + E_{\text{c}}.
\end{align}
Considering only first order transitions and neglecting the nuclear recoil,
the expected spectrum for the excitation energy is characterized by a sum of Breit-Wigner resonances modulated by the phase space factor
(see Refs.~\cite{Drexlin:2013lha,Dragoun:2015oja,Nucciotti:2015rsl}):
\begin{align}
\frac{dn_{\text{EC}}}{dE_{\text{c}}}
\propto
&
(Q-E_{\text{c}}) \sum_{k=1}^{N}
|U_{ek}|^2
\sqrt{(Q-E_{\text{c}})^2 - m_{k}^2}
\nonumber
\\
&
\times
\Theta(Q-E_{\text{c}}-m_{k})
\nonumber
\\
&
\times
\sum_i
P_{i}
\frac{\Gamma_i/2\pi}{(E_{\text{c}}-E_i)^2+\Gamma_i^2/4}
.
\label{lambda}
\end{align}
Here,
$P_{i}$
is the probability of electron capture from the $i$-shell,
which has been calculated in Ref.~\cite{Faessler:2014xpa} using a fully relativistic approach.
It is given by
$
P_{i}
=
|\psi_{i}(R)|^2
B_{i}
$,
where $|\psi_{i}(R)|^2$ is the square of single electron wave functions of the parent atom at the nuclear radius $R$ and $B_i$ is a correction for electron exchange and overlap.
The energy $E_i$ is the peak energy of the $i$-th resonance,
which is given in a first approximation by the difference between the binding energy
in the daughter atom of the electron that has been captured
and the binding energy of the 4$f$ electron: $E_i \simeq E^b_i-E^b_{4f}$.
The width $\Gamma_i$ is the intrinsic width of the resonance,
which is related to the half-life of the excited $i$-state.
The Heaviside function $\Theta(Q-E_{\text{c}}-m_{k})$
ensures the reality of the expression.
The parameters describing the atomic excited states are taken from Ref. \cite{Faessler:2014xpa}
and listed in Tab.~\ref{tab:BW}.

The fraction of the calorimetrically measured spectrum which is mostly affected by finite neutrino masses is the endpoint region, where the emitted neutrino has only a few eV of kinetic energy.
In the following,
we consider a detector with energy resolution
of 5 or 2 eV
and we assume that the masses
$m_{1}$,
$m_{2}$,
$m_{3}$
of the three massive neutrinos
$\nu_{1}$,
$\nu_{2}$,
$\nu_{3}$,
in the framework of the standard three-neutrino mixing scenario,
are much smaller than the energy resolution.
In this case,
Eq.~(\ref{lambda})
can be approximated by
\begin{align}
\left(
\frac{dn_{\text{EC}}}{dE_{\text{c}}}
\right)_{3\nu}
\propto
&
(Q-E_{\text{c}})
\sqrt{(Q-E_{\text{c}})^2 - m_{\nu}^2}
\nonumber
\\
&
\times
\Theta(Q-E_{\text{c}}-m_{\nu})
\nonumber
\\
&
\times
\sum_i
P_{i}
\frac{\Gamma_i/2\pi}{(E_{\text{c}}-E_i)^2+\Gamma_i^2/4}
,
\label{lambda3nu}
\end{align}
with the effective electron neutrino mass
\begin{equation}
m_{\nu}^2
=
\sum_{k=1}^{3}
|U_{ek}|^2 m_{k}^2
\label{mnu}
\end{equation}
This approximation is consistent with the most stringent upper limits on
$m_{\nu}$
found in the Mainz \cite{Kraus:2004zw} and Troitsk \cite{Aseev:2011dq} experiments:
\begin{equation}
m_{\nu} \leq
\left\{
\begin{array}{ll} \displaystyle
2.3 \, \text{eV}
&
\quad
(\text{Mainz})
,
\\ \displaystyle
2.05 \, \text{eV}
&
\quad
(\text{Troitsk})
,
\end{array}
\right.
\label{mb}
\end{equation}
at 95\% CL.

\begin{table}
\centering
\caption{
Experimental excitation energies $E_{i}$ of the hole states with their widths $\Gamma_{i}$
and
$P_{i}/P_{\text{M1}}$.
Data taken from Ref.~\cite{Faessler:2014xpa}.
}
\label{tab:BW}
\begin{tabular}{crdd}
Level $i$ &
$E_i$ (eV) &
\multicolumn{1}{c}{$\Gamma_i$ (eV)} &
\multicolumn{1}{c}{$P_{i}/P_{\text{M1}}$}
\\ \hline
M1 & 2040 & 13.7  & 1     \\
M2 & 1836 &  7.2  & 0.051 \\
N1 &  411 &  5.3  & 0.244 \\
N2 &  333 &  8.0  & 0.012 \\
O1 &   48 &  4.3  & 0.032 \\
\hline
\end{tabular}
\end{table}

\section{The ECHo experiment}
\label{sec:ECHo}

The ECHo experiment is designed to reach a
sub-eV sensitivity to the electron neutrino mass
through the analysis of the endpoint region of the $^{163}$Ho spectrum.
The concept at the basis of this experiment is that all the energy released during the $^{163}$Ho electron capture,
besides that taken away by the  neutrino, is measured with high precision.
Large arrays of low temperature metallic magnetic calorimeters (MMCs) \cite{Fleischmann2005}
will be used.
The $^{163}$Ho atoms will be completely enclosed in the energy absorber,
which consists of a gold film with about 10 $\mu$m thickness
and a $200 \times 200 \, \mu$m$^2$ surface area.
Such an absorber is thermally coupled to a temperature sensor,
which is a thin film of a paramagnetic material,
typically gold doped with a few hundreds ppm of erbium,
sitting in an external stable magnetic field. The sensor is then weakly coupled to the thermal bath kept at a constant temperature of less then 30 mK. When energy is deposited in the detector, its temperature increases leading to a change of magnetization of the sensor which is read out as a change of flux by low-noise high-bandwidth dc-SQUIDs
(Superconducting QUantum Interference Devices).
An energy resolution as good as 1.6 eV FWHM at 6 keV has already been achieved with MMCs developed for soft x-ray spectroscopy as well as very precise calibration functions \cite{Pies:2012nua}.
An intrinsic background is the unresolved pileup which
is related to the finite time resolution of the detector
and to the fact that, since the $^{163}$Ho is enclosed in the detector itself,
each $^{163}$Ho decay leads to a signal.
Therefore,
two or more events which occur in a time interval shorter than the risetime of the pulse
are misidentified as a single event
with an energy given approximately by the sum of the single event energies.
The fraction of pileup events is given by the product of the activity in the detector and the risetime of the signal.
In order to be able to investigate small neutrino masses, the unresolved pileup fraction $f_{\text{pp}}$ should be smaller than $10^{-5}$.
The first prototypes of MMCs with embedded $^{163}$Ho have already shown a risetime of the order of 100 ns \cite{Gastaldo:2012nv},
which allows for single pixel activities of the order of a few tens of Bq.
The goal of the ECHo experiment is to have
the sum of all other background contributions in the endpoint region of the spectrum
at least one order of magnitude smaller than the unresolved pileup.
This corresponds to a background parameter
$b < 5 \times 10^{-5}$ counts/eV/det/day.

During the first phase of the ECHo experiment, ECHo-1k,
which already started,
more then $10^{10}$ events
of $^{163}$Ho electron capture will be collected
in one year of measuring time
by having a $^{163}$Ho source of the order of 1000 Bq distributed into about 100 MMCs.
The major goals of this phase are to obtain an energy resolution better than 5 eV FWHM for multiplexed detectors and an unresolved pileup fraction smaller than $10^{-5}$.
Achieving these goals will allow the ECHo Collaboration
to reach a limit on the electron neutrino mass below 10 eV,
which is more than one order of magnitude better than the current
limit on the electron neutrino mass
obtained with a
$^{163}$Ho electron capture experiment, $m_{\nu} < 225$ eV at 95\% C.L. \cite{Springer:1987zz}.

In the second phase of ECHo, called ECHo-1M,
a $^{163}$Ho source of the order of 1 MBq will be embedded in a large number of pixels divided into multiplexed arrays. The aim of this phase is to measure a $^{163}$Ho spectrum with about $10^{14}$ events with an energy resolution better that 2 eV FWHM and an unresolved pileup fraction of the order of $10^{-6}$. With ECHo-1M the sensitivity to the electron neutrino mass will reach the sub-eV region \cite{ECHo-2016}.

The discussed sensitivities are based on the analysis of simulated $^{163}$Ho spectra which are generated using only the first order excited states in $^{163}$Dy. Higher order excited states, like the one corresponding to the formation of two holes in the $^{163}$Dy atom after the electron capture, even if they have a much smaller probability to occur, can play a quite important role in the region near the endpoint of the spectrum.
The role of higher order excitations has been recently studied in Refs.~\cite{Robertson:2014fka,Faessler:2015pka,DeRujula:2015lya,DeRujula:2016fdu}.
There is still not a good agreement among the different authors on the expected structures in the $^{163}$Ho spectrum due to these excitations.
The available data on the $^{163}$Ho spectrum
\cite{Ranitzsch:2012nua,Ranitzsch:2014kma,Croce:2015kwa}
are still not able to clearly resolve the controversy.
An important point to mention is that the two-hole excitations
in which an electron is ``shaken-off'' in the continuum
may imply a substantial increase of the fraction of events in the endpoint region of the spectrum
\cite{DeRujula:2015lya,DeRujula:2016fdu}.
Therefore, by presenting limits on the sensitivity based only on the first order excited states,
we provide upper values of the sensitivity that could be reached with a well-defined experimental configuration.

\section{$3\nu$ mixing}
\label{sec:3nu}

In this section we describe our methodology to obtain the sensitivity for the neutrino mass in the ECHo experiment
and we present our results for the sensitivity to $m_{\nu}$
in the standard case of three-neutrino mixing.
Previous analyses of the sensitivity of $^{163}$Ho experiments
with various configurations have been presented in
Refs.~\cite{Nucciotti:2009wq,Gatti:2012ii,Nucciotti:2014raa,Gastaldo:2014hga}.

The theoretical spectrum of $^{163}$Ho electron capture events as a function of
the total released energy $E_{\text{c}}$ is given by
\begin{equation}
\frac{dn}{dE_{\text{c}}}(m_{\nu})
=
N_{\text{ev}}
S_{\text{tot}}(E_{\text{c}},m_{\nu})
\otimes
R_{\Delta{E}}(E_{\text{c}})
+ B
,
\label{spectrum}
\end{equation}
with the normalized total spectrum
\begin{align}
S_{\text{tot}}(E_{\text{c}},
\null & \null
m_{\nu})
=
\left(1+f_{\text{pp}}\right)^{-1}
\big[
S_{\text{EC}}(E_{\text{c}},m_{\nu})
\nonumber
\\
\null & \null
+
f_{\text{pp}}
S_{\text{EC}}(E_{\text{c}},m_{\nu})
\otimes
S_{\text{EC}}(E_{\text{c}},m_{\nu})
\big]
.
\label{stot}
\end{align}
Here $S_{\text{EC}}(E_{\text{c}},m_{\nu})$ is the normalized electron-capture spectrum
\begin{equation}
S_{\text{EC}}(E_{\text{c}},m_{\nu})
=
\left(
\frac{dn_{\text{EC}}}{dE_{\text{c}}}
\right)_{\hspace{-0.1cm}3\nu}
\left(
\int_{0}^{Q-m_{\nu}}
\left(
\frac{dn_{\text{EC}}}{dE_{\text{c}}}
\right)_{\hspace{-0.1cm}3\nu}
\hspace{-0.1cm}
d E_{\text{c}}
\right)^{-1}
,
\label{NEC0}
\end{equation}
with $dn_{\text{EC}}/dE_{\text{c}}$ given by Eq.~(\ref{lambda}).
Other quantities in Eqs.~(\ref{spectrum}) and (\ref{stot}) are:
the total number of events
$N_{\text{ev}}$,
which in a real experiment is given by
$N_{\text{ev}} = N_{\text{det}} A t_{\text{m}}$,
where $N_{\text{det}}$ is the number of detectors,
$A$ is the activity of the $^{163}$Ho source in each detector
and $t_{\text{m}}$ is the measuring time;
the background\footnote{
For simplicity,
we assume an energy-independent background.
If the background has an energy dependence
it must be included in the convolution with the energy resolution.
}
$B = b t_{\text{m}}$;
the fraction of pileup events $f_{\text{pp}}$,
that, in a first approximation,
is given by $f_{\text{pp}} = \tau_{\text{R}} A$,
where $\tau_{\text{R}}$ is the time resolution.
The detector energy response
$R_{\Delta{E}}(E_{\text{c}})$
is assumed to be Gaussian:
\begin{equation}
R_{\Delta{E}}(E_{\text{c}}) = \frac{1}{\sigma_{\Delta{E}} \sqrt{2\pi}}\exp{(-E_{\text{c}}^2/2\sigma_{\Delta{E}}^2)},
\end{equation}
with variance relate to the full width at half maximum by the usual relation
$\sigma_{\Delta{E}} = \Delta{E}_{\text{FWHM}}/2.35$.
In Eqs.~(\ref{spectrum}) and (\ref{stot}),
the symbol $\otimes$ represents a convolution.
The self-convolution of the normalized spectrum in the second term of Eq.~(\ref{stot})
accounts for the pileup effect.
In order to speed up the computer-intensive evaluation
of the sensitivity to $m_{\nu}$,
in this term
we used the normalized spectrum
$S_{\text{EC}}(E_{\text{c}},0)$,
neglecting the small effects due to $m_{\nu}$.

\begin{figure}
\centering
\includegraphics[width=\linewidth]{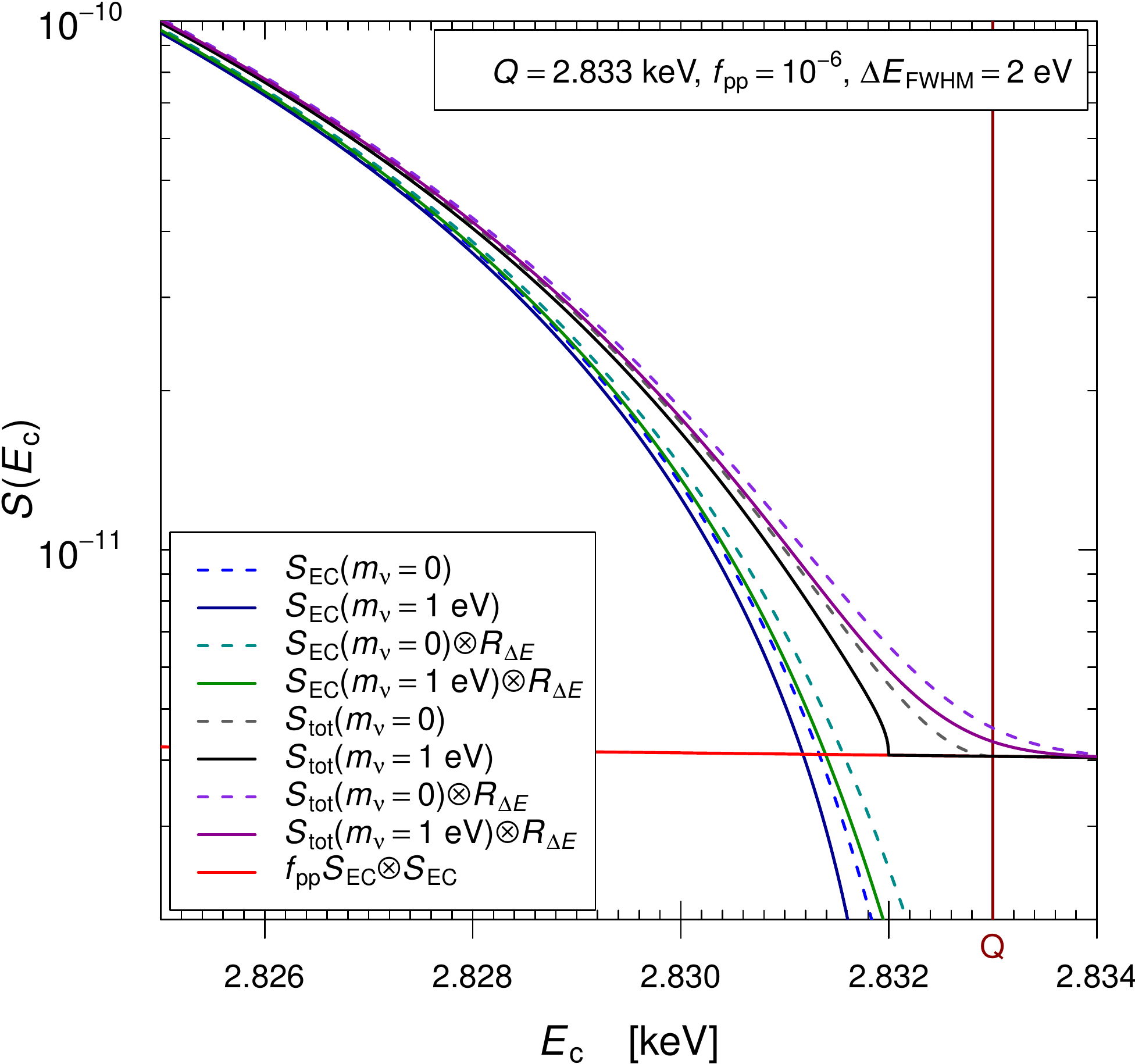}
\caption{Energy spectra calculated without and with the convolution with the
detector energy response
$R_{\Delta{E}}(E_{\text{c}})$
for $m_{\nu} = 0$ and for $m_{\nu} = 1 \, \text{eV}$.
}
\label{fig:spectrum}
\end{figure}

Figure~\ref{fig:spectrum}
illustrates the effect of an effective neutrino mass
$m_{\nu} = 1 \, \text{eV}$
on the
spectrum
$S_{\text{EC}}$
and on the total spectrum
$S_{\text{tot}}$
without and with the convolution with the
detector energy response
$R_{\Delta{E}}(E_{\text{c}})$
for
$\Delta{E}_{\text{FWHM}} = 2 \, \text{eV}$.
One can see that in the limit of negligible unresolved pileup,
represented by the curves labeled $S_{\text{EC}}$,
the difference between the spectra with
$m_{\nu} = 0$
and
$m_{\nu} = 1 \, \text{eV}$
without and with
the convolution with the
detector energy response
is similar.
On the other hand,
the difference of the total spectra
$S_{\text{tot}}$
for
$m_{\nu} = 0$
and
$m_{\nu} = 1 \, \text{eV}$
is significantly affected by the energy resolution of the detector.
Without considering the finite energy resolution of the detector,
the difference between
$S_{\text{tot}}(m_{\nu} = 0)$
and
$S_{\text{tot}}(m_{\nu} = 1 \, \text{eV})$
is relatively large
around
$Q - m_{\nu}$,
where
$S_{\text{EC}}(m_{\nu} = 1 \, \text{eV})$
vanishes and only the pileup
contributes.
Since
this difference is strongly reduced by
the convolution with the
detector energy response,
it is clear that the sensitivity to the neutrino mass
depends on the
energy resolution of the detector.
However,
the effects of
a poor energy resolution can be counterbalanced by a large statistics $N_{\text{ev}}$
which allows to distinguish the difference between
$dn/dE_{\text{c}}(m_{\nu}\neq0)$ and $dn/dE_{\text{c}}(m_{\nu}=0)$.
Indeed,
since the difference is proportional to $N_{\text{ev}}$,
the Poisson fluctuations of the event numbers in the energy bins are proportional to
$\sqrt{N_{\text{ev}}}$
and the sensitivity to
$m_{\nu}^2$ is proportional to $N_{\text{ev}}^{-1/2}$,
leading to a sensitivity to
$m_{\nu}$ proportional to $N_{\text{ev}}^{-1/4}$
(see also the discussions in
Refs.~\cite{Nucciotti:2009wq,Nucciotti:2015rsl}).

\begin{figure}
\centering
\includegraphics[width=\linewidth]{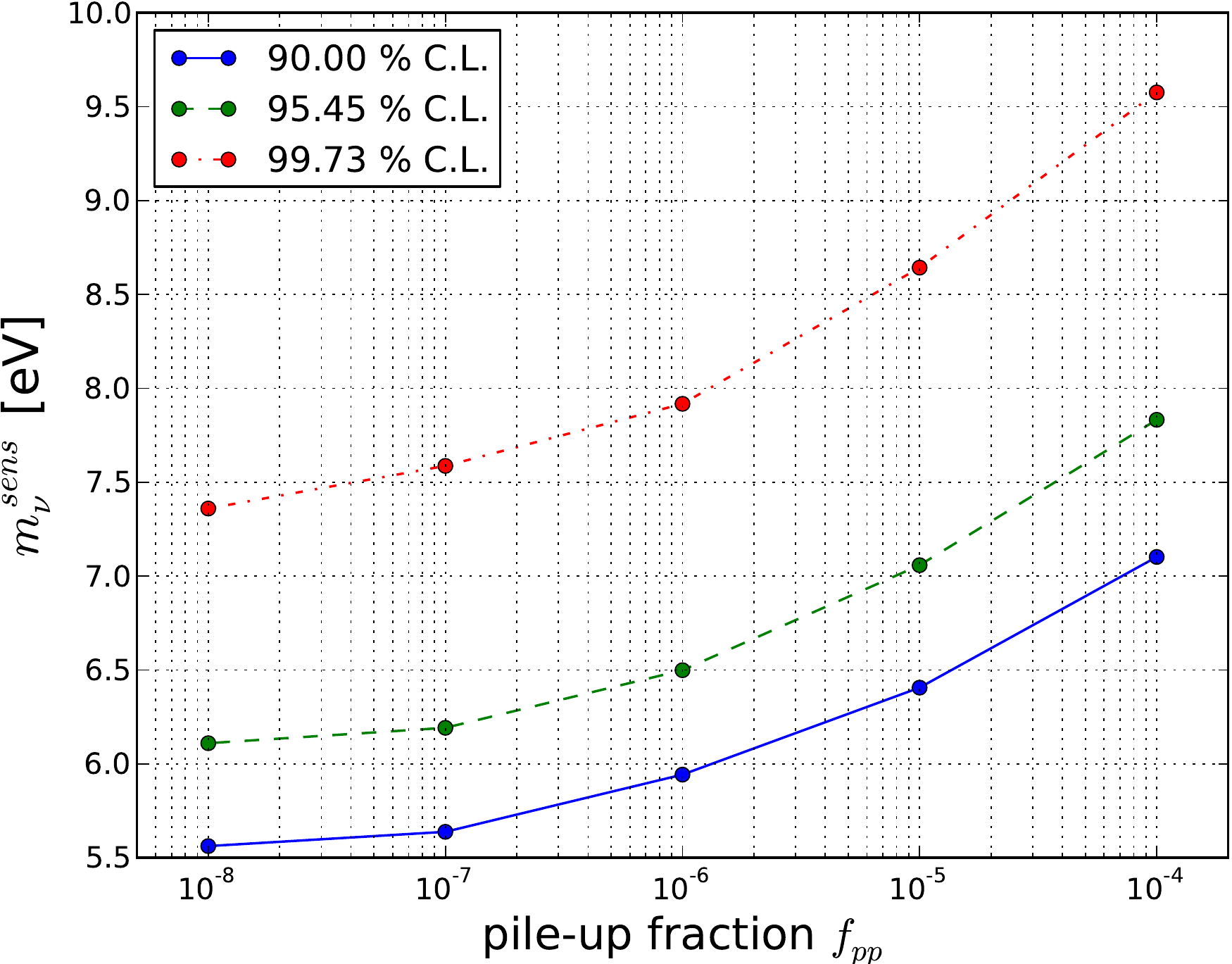}
\caption{Estimated sensitivity to $m_{\nu}$ in the ECHo-1k experiment as a function of the pileup fraction $f_{\text{pp}}$.
We used
$N_{\text{sim}} = 1000$
simulations generated with
$N_{\text{ev}} = 10^{10}$,
$Q = 2.833$ keV,
$\Delta{E}_{\text{FWHM}}= 5$ eV
and $B = 0$.}
\label{fig:sensitivityfppN10E5}
\end{figure}

\begin{figure}
\centering
\includegraphics[width=\linewidth]{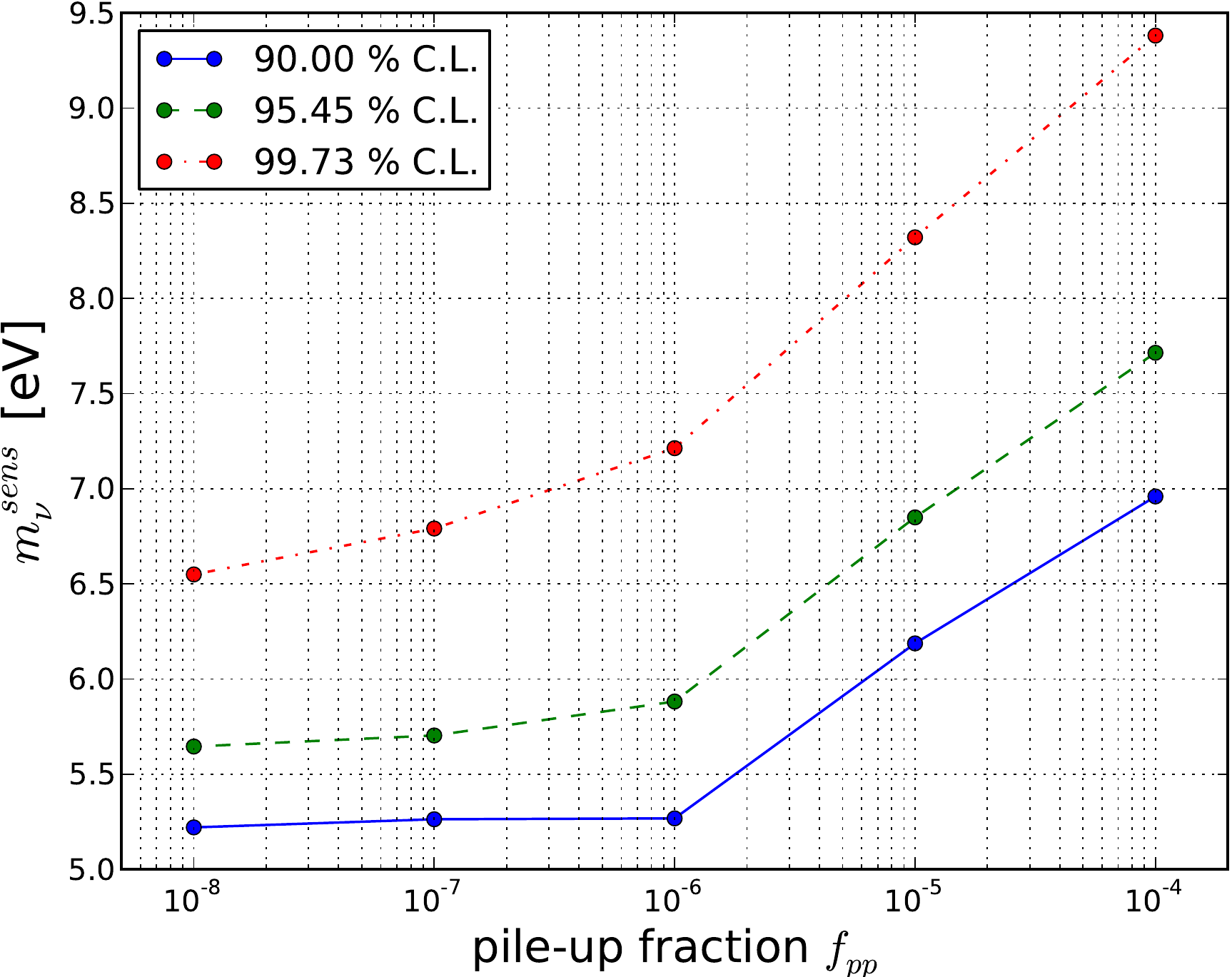}
\caption{Estimated sensitivity to $m_{\nu}$ as a function of the pileup fraction $f_{\text{pp}}$
in the beginning of the the ECHo-1M experiment
when the same statistics of $N_{\text{ev}} = 10^{10}$
expected in the ECHo-1k
will be reached.
We used
$N_{\text{sim}} = 1000$
simulations generated with
$Q = 2.833$ keV,
$\Delta{E}_{\text{FWHM}}= 2$ eV
and $B = 0$.}
\label{fig:sensitivityfppN10E2}
\end{figure}

\begin{figure}
\centering
\includegraphics[width=\linewidth]{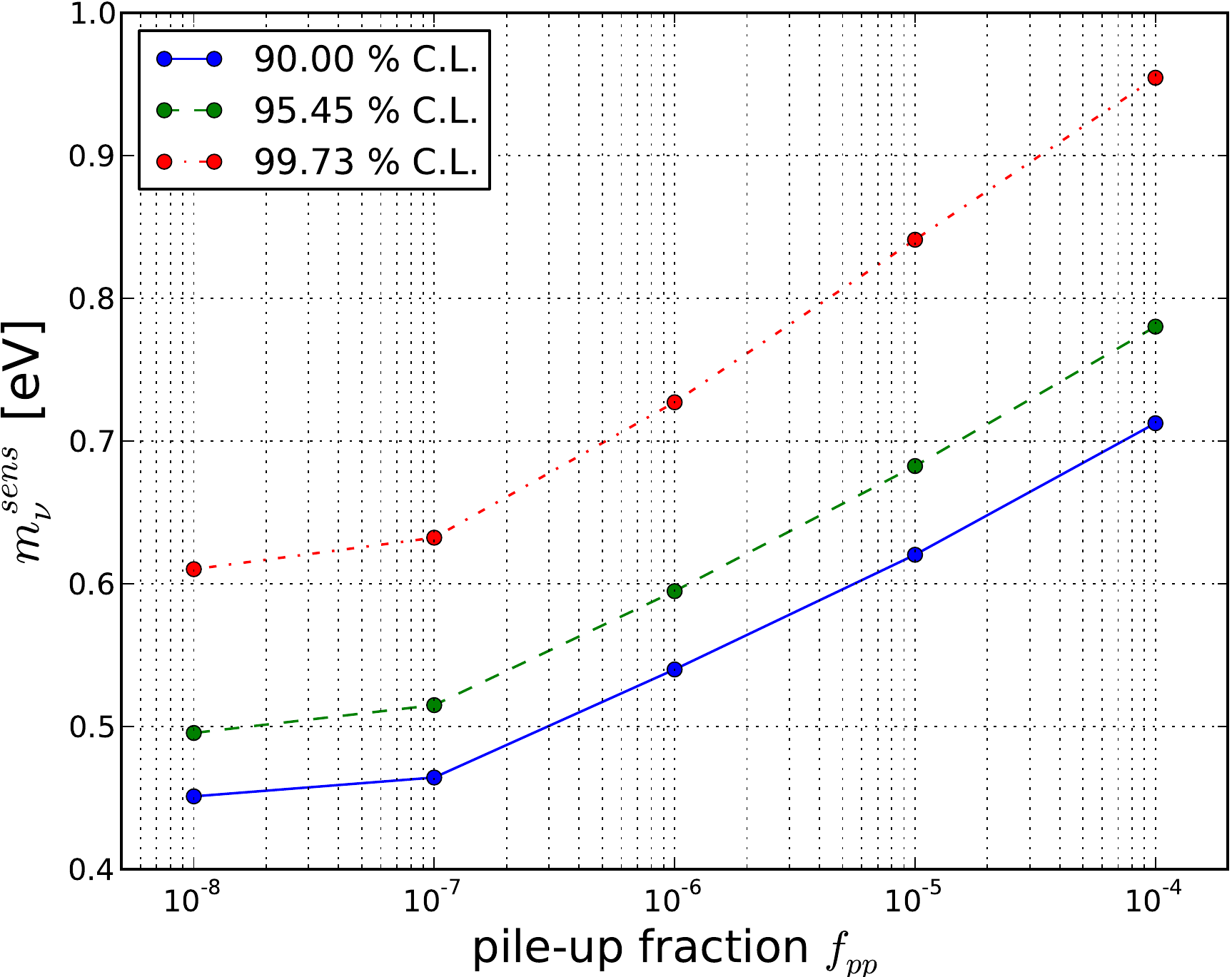}
\caption{Estimated sensitivity to $m_{\nu}$ in the ECHo-1M experiment as a function of the pileup fraction $f_{\text{pp}}$.
We used
$N_{\text{sim}} = 1000$
simulations generated with
$N_{\text{ev}} = 10^{14}$,
$Q = 2.833$ keV,
$\Delta{E}_{\text{FWHM}}= 2$ eV
and $B = 0$.}
\label{fig:sensitivityfppN14E2}
\end{figure}

We computed the sensitivity
$m_{\nu}^{\text{sens}}$
to $m_{\nu}$
of a given experimental configuration
defined by the energy resolution of the detectors,
the unresolved pileup fraction and the total statistics.
We adopted the Feldman-Cousins definition of sensitivity\footnote{
Note that our definition of sensitivity is different of that used
in Refs.~\cite{Nucciotti:2009wq,Gatti:2012ii,Nucciotti:2014raa}.
}
given in Ref.~\cite{Feldman:1997qc}:
``the sensitivity is defined as the average upper
limit one would get from an ensemble of experiments with
the expected background and no true signal.''
Hence,
for a given experimental configuration
we generated $N_{\text{sim}}$ simulations of the data in the case $m_{\nu}=0$,
for each simulation we found the corresponding upper limit for $m_{\nu}$,
and we calculated the sensitivity as the median of these upper limits.
We did not use the mean of the upper limits,
which may be interpreted as the ``average'' in the
Feldman-Cousins definition of sensitivity,
because the mean is not defined in the case of
limits on more than one parameter,
as in the case of 3+1 neutrino mixing considered in Section~\ref{sec:3p1}.
On the other hand,
for $N_{\text{par}}$ parameters
the median is defined as the $N_{\text{par}}$ hypersurface
which encloses all the values of the parameters
which are allowed by more than 50\% of the simulations\footnote{
Note, however, that
in the one-parameter case the distinction is
practically irrelevant if the fluctuations of the simulations
follow a Gaussian distribution,
for which the mean is equal to the median.
In our case we use a Poisson distribution,
but since the number of events in the bins are large if the pileup is not too small,
the distinction between median and mean is negligible in our analysis.
}.

We considered two experimental configurations corresponding
to the expected performances of the
ECHo-1k and ECHo-1M experiments
\cite{Gastaldo:2013wha,Hassel:2016echo}.
For ECHo-1k we considered
$\Delta{E}_{\text{FWHM}} = 5 \, \text{eV}$
and
$N_{\text{ev}} = 10^{10}$,
whereas for ECHo-1M we considered
$\Delta{E}_{\text{FWHM}} = 2 \, \text{eV}$
and
$N_{\text{ev}} = 10^{14}$.
We considered different values of the pileup fraction $f_{\text{pp}}$
from $10^{-8}$ to $10^{-4}$.
We also neglected the background $B$,
which in the ECHo experiment is expected to be
at least one order of magnitude smaller than the unresolved pileup,
as already mentioned above
(see also the discussion in Ref.~\cite{Nucciotti:2014raa}).

The simulations have been generated with
$Q = 2.833 \, \text{keV}$
and
the simulated data have been fitted from
$E_{\text{c}}^{\text{min}} = 2.2 \, \text{keV}$
to
$E_{\text{c}}^{\text{max}} = 3.2 \, \text{keV}$
with different bin sizes.
We checked that the results are
independent of the bin size as long as it is smaller than the
energy resolution uncertainty $\sigma_{\Delta{E}}$.

The theoretical average number of events
in the $i^{\text{th}}$ energy bin
(with $i=1,\ldots,N_{\text{bins}}$)
is given by
\begin{equation}
n^{\text{th}}_{i}(m_{\nu})
=
\int_{E_{i}^{\text{min}}}^{E_{i}^{\text{max}}}
\frac{dn}{dE_{\text{c}}}(m_{\nu})
\,
d E_{\text{c}}
,
\label{numberevents}
\end{equation}
where
$E_{i}^{\text{min}}$ and $E_{i}^{\text{max}}$
are, respectively, the lower and upper borders of the bin.
In the $j^{\text{th}}$ simulation of the data (with $j=1,\ldots,N_{\text{sim}}$),
the number of events
$(n^{\text{sim}}_{i})_{j}$
in the $i^{\text{th}}$ bin
is obtained with a Poisson fluctuation
around the theoretical average number of events $n^{\text{th}}_{i}(0)$,
corresponding to $m_{\nu}=0$.
The $\chi^2$ of the $j^{\text{th}}$ simulation is given by
\begin{align}
\chi^2_j(m_{\nu})
=
&
2 \sum_{i=1}^{N_{\text{bins}}}
n^{\text{th}}_{i}(m_{\nu}) - (n^{\text{sim}}_{i})_{j}
\nonumber
\\
&
+
(n^{\text{sim}}_{i})_{j}
\ln\!\left( \frac{(n^{\text{sim}}_{i})_{j}}{n^{\text{th}}_{i}(m_{\nu})} \right)
.
\end{align}
Although specific values of $Q$, $N_{\text{ev}}$, $f_{\text{pp}}$ and $B$
have to be used for the generation of the simulated $(n^{\text{sim}}_{i})_{j}$,
we do not make any assumption for the values of these parameters
in the expression of $n^{\text{th}}_{i}(m_{\nu})$
used in the fit of the simulated data
and $\chi^2_j(m_{\nu})$ is calculated by marginalizing over them.
This method reflects the probable real experimental approach,
in which these parameters will be determined by the data\footnote{
We kept fixed the energy and width of the M1 Breit-Wigner resonance
whose tail determines the spectrum
in the energy range of the fits.
These parameters will be measured independently with high precision in
ECHo and other $^{163}$Ho experiments.
}.

For each simulation $j$ we compute the
upper limit
$(m_{\nu}^{\text{UL}})_{j}$
for $m_{\nu}$
at $CL$ confidence level
using the relation:
\begin{equation}
\chi^2_j((m_{\nu}^{\text{UL}})_{j})
=
(\chi^2_j)_{\text{min}}
+
\Delta\chi^2(CL)
,
\label{deltachi3nu}
\end{equation}
where
$(\chi^2_j)_{\text{min}}$
is the minimum of
$\chi^2_j(m_{\nu})$
and
$
\Delta\chi^2(CL)
=
2.71,
4.0,
9.0
$
for
$
CL
=
90\%,
95.45\%,
99.73\%
$,
respectively.
As explained above,
the sensitivity
$m_{\nu}^{\text{sens}}$
is given by the median of the upper limits $(m_{\nu}^{\text{UL}})_{j}$
in the ensemble of $N_{\text{sim}}$ simulations.

For the first stage of the ECHo experiment, ECHo-1k,
the aim is to achieve a total statistics of $N_{\text{ev}} \simeq 10^{10}$
with an energy
resolution $\Delta{E}_{\text{FWHM}} \simeq 5$ eV.
Figure~\ref{fig:sensitivityfppN10E5}
shows our estimation of the sensitivity to $m_{\nu}$
of ECHo-1k as a function of $f_{\text{pp}}$.
One can see that for the foreseen value $f_{\text{pp}} \simeq 10^{-6}$
the sensitivity will be around 6.5 (7.9) eV at $2\sigma$ ($3\sigma$),
which will represent an improvement of
more than one order of magnitude with respect to
the current limit
$m_{\nu} < 225$ eV at $2\sigma$ \cite{Springer:1987zz}
obtained with a
$^{163}$Ho electron capture experiment.
One can also notice that
the sensitivity does not improve much
decreasing the value of $f_{\text{pp}}$
below about $10^{-6}$.
This happens for the following two reasons:

\begin{enumerate}

\item
\label{c1}
The relative contribution of the pileup to the number of events
is negligible
in an energy interval of the order of the energy resolution
$\Delta{E}_{\text{FWHM}}$
near the endpoint.
Indeed,
near the endpoint
$S_{\text{EC}} \propto \Delta{E}_{\text{FWHM}}^2/Q^3$
and the number of events in the energy interval $\Delta{E}_{\text{FWHM}}$ is proportional to
$(\Delta{E}_{\text{FWHM}}/Q)^3$.
On the other hand,
since
typically the pileup is due to two events with energies well below the endpoint,
where
$Q-E_{\text{c}}$
is large,
the number of pileup events in the energy interval $\Delta{E}_{\text{FWHM}}$ is proportional to
$f_{\text{pp}}\Delta{E}_{\text{FWHM}}/2Q$.
Hence,
the pileup is negligible near the endpoint for
$f_{\text{pp}} \ll 2 (\Delta{E}_{\text{FWHM}}/Q)^2$,
i.e.
$f_{\text{pp}} \ll 5 \times 10^{-6}$
for
$\Delta{E}_{\text{FWHM}} \simeq 5$ eV.

\item
\label{c2}
The average number of pileup events
in an energy interval of the order of the energy resolution
$\Delta{E}_{\text{FWHM}}$
near the endpoint
is smaller than one.
Indeed,
neglecting the small effects due to the neutrino mass,
the average number of pileup events
in the energy interval
$\Delta{E}_{\text{FWHM}}$
is smaller than one for
\begin{equation}
f_{\text{pp}}
\lesssim
\left[
N_{\text{ev}}
S_{\text{EC}}(E_{\text{c}},0)
\otimes
S_{\text{EC}}(E_{\text{c}},0)
\Delta{E}_{\text{FWHM}}
\right]^{-1}
.
\label{fpp1}
\end{equation}
Since near the endpoint
we have
$
S_{\text{EC}}(E_{\text{c}},0)
\otimes
S_{\text{EC}}(E_{\text{c}},0)
=
4.07 \times 10^{-6}
$,
for
$N_{\text{ev}}=10^{10}$
and
$\Delta{E}_{\text{FWHM}} \simeq 5$ eV
we obtain the condition
$
f_{\text{pp}}
\lesssim
5 \times 10^{-7}
$.

\end{enumerate}

In the second stage of the ECHo experiment, ECHo-1M, it is expected to have
an energy resolution better than $\Delta{E}_{\text{FWHM}} = 2$ eV.
Figure~\ref{fig:sensitivityfppN10E2}
shows our estimation of the sensitivity to $m_{\nu}$
of ECHo-1M as a function of $f_{\text{pp}}$
when the same statistics of $N_{\text{ev}} = 10^{10}$
expected in the ECHo-1k
will be reached.
Comparing Figs.~\ref{fig:sensitivityfppN10E5} and \ref{fig:sensitivityfppN10E2},
one can see that the improvement of the energy resolution
generates a small improvement of the sensitivity.
One can also notice a flatter behavior of the
sensitivity for $f_{\text{pp}} \lesssim 10^{-6}$
in Fig.~\ref{fig:sensitivityfppN10E2}
than in Fig.~\ref{fig:sensitivityfppN10E5}.
This is due to the fact that albeit the condition
\ref{c1} above
is satisfied for
$f_{\text{pp}} \ll 1 \times 10^{-6}$,
the condition \ref{c2}
is already satisfied for
$f_{\text{pp}} \lesssim 1 \times 10^{-6}$.

Figure~\ref{fig:sensitivityfppN14E2}
shows our estimation of the final sensitivity to $m_{\nu}$
of ECHo-1M as a function of $f_{\text{pp}}$
when the statistics of $N_{\text{ev}} = 10^{14}$
will be reached.
One can see that it is possible to reach a sensitivity of about 0.6 (0.7) eV
at $2\sigma$ ($3\sigma$)
for the foreseen value $f_{\text{pp}} \simeq 10^{-6}$.
Hence,
ECHo-1M will enter into the sub-eV region of $m_{\nu}$,
not far from the expected 0.2 eV sensitivity of
KATRIN
\cite{Osipowicz:2001sq,Mertens-TAUP2015}.
The behavior of the
sensitivity for $f_{\text{pp}} \lesssim 10^{-6}$
is less flat than those
in Fig.~\ref{fig:sensitivityfppN10E5}
and \ref{fig:sensitivityfppN10E2}
because only the condition
\ref{c1} above
is satisfied for
$f_{\text{pp}} \ll 1 \times 10^{-6}$,
whereas the condition \ref{c2}
is satisfied only for
$f_{\text{pp}} \lesssim 1 \times 10^{-10}$.

\begin{figure}
\centering
\includegraphics[width=\linewidth]{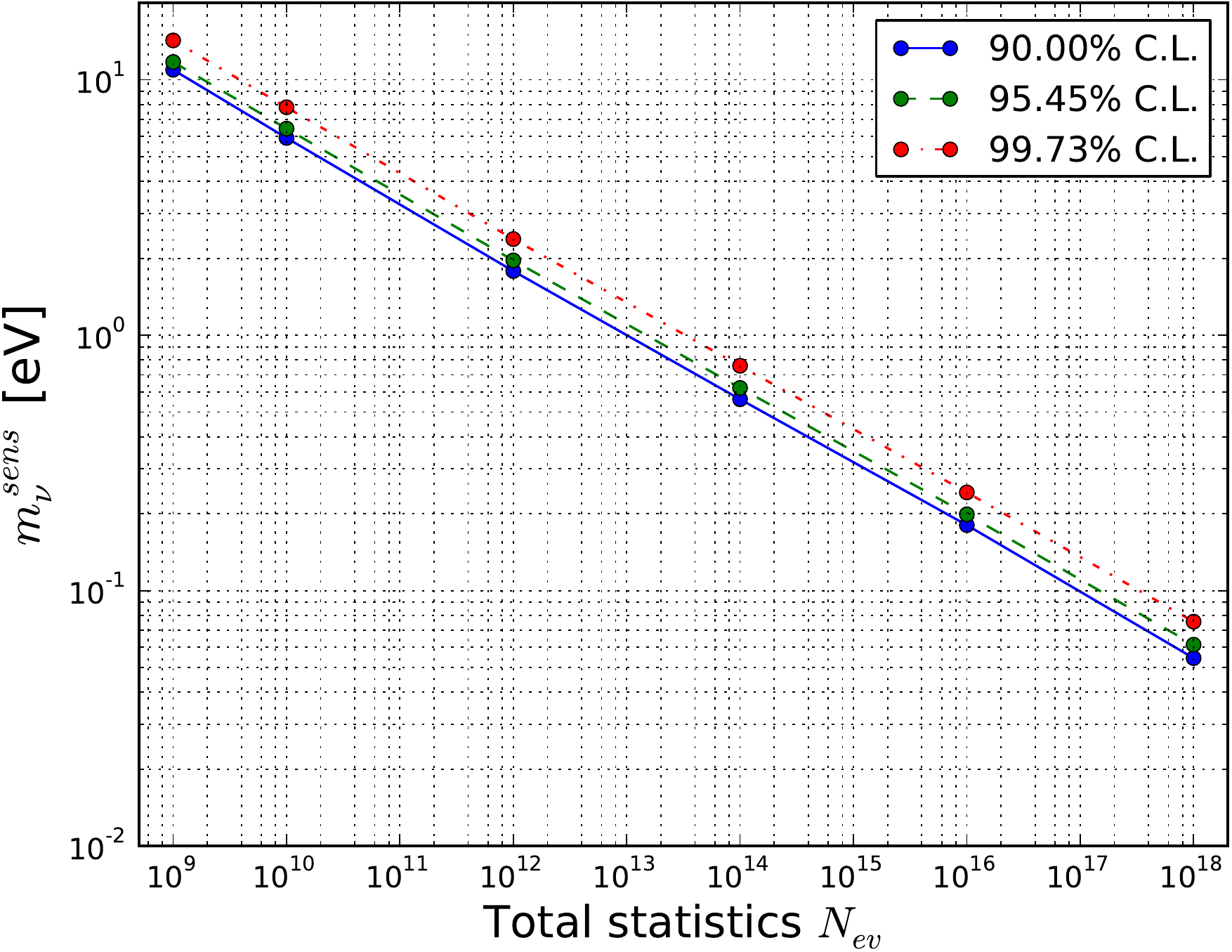}
\caption{Estimated sensitivity to $m_{\nu}$ as a function of the statistics $N_{\text{ev}}$.
We used
$N_{\text{sim}} = 1000$
simulations generated with
$Q = 2.833$ keV,
$\Delta{E}_{\text{FWHM}}= 2$ eV,
$f_{\text{pp}} = 10^{-6}$ and $B = 0$.}
\label{fig:sensitivityNev}
\end{figure}

Figure~\ref{fig:sensitivityNev}
shows our results for the sensitivity to
$m_{\nu}$
as a function of the total statistics $N_{\text{ev}}$
for
$\Delta{E}_{\text{FWHM}}= 2$ eV,
$f_{\text{pp}} = 10^{-6}$ and $B = 0$.
One can see that
$m_{\nu}^{\text{sens}}$ follows the expected proportionality to $N_{\text{ev}}^{-1/4}$
explained above,
in agreement with the calculations presented in
Refs.~\cite{Nucciotti:2014raa,Nucciotti:2015rsl}.

In a future experiment larger than
ECHo-1M
it may be possible to have a total statistics of $N_{\text{ev}} \simeq 10^{16}$.
Figure~\ref{fig:sensitivityNev} shows that in this case it will be possible to reach a sensitivity
to $m_{\nu}$ of
about 0.2 eV,
similar to that expected for the KATRIN experiment
\cite{Osipowicz:2001sq,Mertens-TAUP2015}.

\section{3+1 neutrino mixing}
\label{sec:3p1}

In this section we present our analysis of the sensitivity of future
$^{163}$Ho experiments to the effects of the heavy
neutrino $\nu_{4}$
in the 3+1 neutrino mixing scheme considering
$m_{4} \gg m_{k}$ for $k=1,2,3$
as explained in the introductory Section~\ref{sec:intro}.
In this case,
Eq.~(\ref{lambda})
can be approximated by
\begin{align}
&
\left(
\frac{dn_{\text{EC}}}{dE_{\text{c}}}
\right)_{3+1}
\propto
(Q-E_{\text{c}})
\sum_i
P_{i}
\frac{\Gamma_i/2\pi}{(E_{\text{c}}-E_i)^2+\Gamma_i^2/4}
\nonumber
\\
&
\times
\Big[
(1-|U_{e4}|^2)
\sqrt{(Q-E_{\text{c}})^2 - m_{\nu}^2}
\,
\Theta(Q-E_{\text{c}}-m_{\nu})
\nonumber
\\
&
+
|U_{e4}|^2
\sqrt{(Q-E_{\text{c}})^2 - m_{4}^2}
\,
\Theta(Q-E_{\text{c}}-m_{4})
\Big]
,
\label{lambdasterile1}
\end{align}
with $m_{\nu}$ given by Eq.~(\ref{mnu}).
Therefore,
the complete spectrum can be described as a sum of two spectra, one ending at $Q-m_{\nu}$ with a fraction of events given by $(1-|U_{e4}^2|)$ and the other ending at $Q-m_{4}$ with a fraction of events given by $|U_{e4}^2|$.

The spectrum in Eq.~(\ref{lambdasterile1}) depends on the three neutrino parameters
$m_{\nu}$,
$m_{4}$
and
$|U_{e4}|^2$
and allows to calculate the sensitivity of a $^{163}$Ho in the corresponding
three-dimensional parameter space.
Here, we simplify the problem by assuming
that $m_{\nu}$ is much smaller than the sensitivity of the experiment.
Hence,
we consider the simplified spectrum
\begin{align}
&
\left(
\frac{dn_{\text{EC}}}{dE_{\text{c}}}
\right)_{3+1}
\propto
(Q-E_{\text{c}})
\sum_i
P_{i}
\frac{\Gamma_i/2\pi}{(E_{\text{c}}-E_i)^2+\Gamma_i^2/4}
\nonumber
\\
&
\times
\Big[
(1-|U_{e4}|^2)
(Q-E_{\text{c}})
\,
\Theta(Q-E_{\text{c}})
\nonumber
\\
&
+
|U_{e4}|^2
\sqrt{(Q-E_{\text{c}})^2 - m_{4}^2}
\,
\Theta(Q-E_{\text{c}}-m_{4})
\Big]
,
\label{lambdasterile2}
\end{align}
which depends only on $m_{4}$
and
$|U_{e4}|^2$.

We considered the space of the two parameters
$\Delta{m}^2_{41} \simeq m_{4}^2$
and
$\sin^2 2\vartheta_{ee} = 4 |U_{e4}|^2 ( 1 - |U_{e4}|^2 )$
in order to compare the sensitivity of $^{163}$Ho experiments
with the results of global analyses of short-baseline neutrino oscillation data
\cite{Kopp:2011qd,Giunti:2011gz,Giunti:2011hn,Giunti:2011cp,Conrad:2012qt,Archidiacono:2012ri,Archidiacono:2013xxa,Kopp:2013vaa,Giunti:2013aea,Gariazzo:2015rra,Giunti:2015mwa,Collin:2016rao,Ericson:2016yjn}.
We calculated the sensitivity of $^{163}$Ho experiments in the
$\sin^2 2\vartheta_{ee}$--$\Delta{m}^2_{41}$
plane with a method similar to that described in Section~\ref{sec:3nu},
using the spectrum in Eq.~(\ref{lambdasterile2}).
In the 3+1 case,
for each simulation $j$ we compute the allowed region
at $CL$ confidence level
in the
$\sin^2 2\vartheta_{ee}$--$\Delta{m}^2_{41}$
plane
using the relation:
\begin{equation}
\chi^2_j(\sin^2 2\vartheta_{ee},\Delta{m}^2_{41})
\leq
(\chi^2_j)_{\text{min}}
+
\Delta\chi^2(CL)
,
\label{deltachi3p1}
\end{equation}
where
$(\chi^2_j)_{\text{min}}$
is the minimum of
$\chi^2_j(\sin^2 2\vartheta_{ee},\Delta{m}^2_{41})$
and
$
\Delta\chi^2(CL)
=
4.61,
6.18,
11.83
$
for
$
CL
=
90\%,
95.45\%,
99.73\%
$,
respectively.
We calculate the region of sensitivity
in the
$\sin^2 2\vartheta_{ee}$--$\Delta{m}^2_{41}$
plane
as the set of points which are not allowed by the inequality (\ref{deltachi3p1})
in at least 50\% of the simulations
(see the discussion on the definition of sensitivity
in Section~\ref{sec:3nu}).

\begin{figure}
\centering
\includegraphics[width=\linewidth]{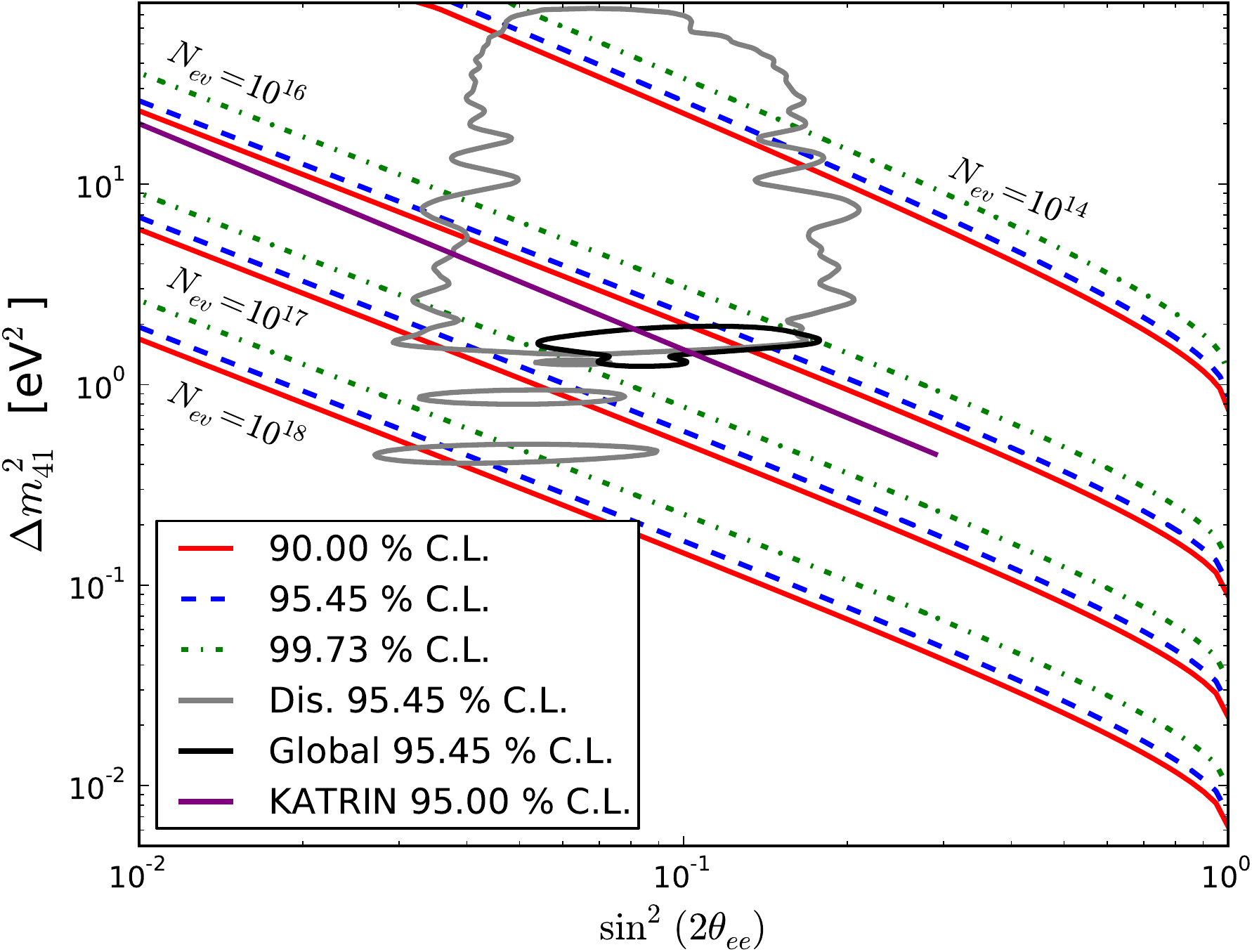}
\caption{Estimated sensitivity curves
at 90\% C.L. (red), 95.45\% C.L. (dashed blue) and 99.73\% C.L. (dash-dotted green)
in the
$\sin^2 2\vartheta_{ee}$--$\Delta{m}^2_{41}$
plane in the case of 3+1 neutrino mixing for
$N_{\text{ev}} = 10^{14}$, $10^{16}$, $10^{17}$ and $10^{18}$.
We used
$N_{\text{sim}} = 100$
simulations generated with
$Q = 2.833$ keV,
$\Delta{E}_{\text{FWHM}}= 2$ eV,
$f_{\text{pp}} = 10^{-6}$
and $B = 0$.
The black curve encloses the region allowed at
95.45\% C.L.
by a global fit of short-baseline neutrino oscillation data \cite{Giunti:2013aea,Gariazzo:2015rra}.
The gray curves enclose the
95.45\% C.L.
allowed regions
obtained by restricting the analysis to the data of $\nu_{e}$ and $\bar\nu_{e}$ disappearance experiments
\cite{Giunti:2012tn,Giunti:2012bc},
taking into account the
Mainz
\cite{Kraus:2012he}
and
Troitsk
\cite{Belesev:2012hx,Belesev:2013cba}
bounds.
Also shown is the expected 95\% C.L. sensitivity of the KATRIN experiment
\cite{Mertens-TAUP2015}.
}
\label{fig:sensitivitysterile}
\end{figure}

The results are presented in Fig.~\ref{fig:sensitivitysterile},
where we plotted the sensitivity curves for
$N_{\text{ev}} = 10^{14}$, $10^{16}$, $10^{17}$ and $10^{18}$,
considering
$Q = 2.833$ keV,
$\Delta{E}_{\text{FWHM}}= 2$ eV and
$f_{\text{pp}} = 10^{-6}$.
From Fig.~\ref{fig:sensitivitysterile} one can see that the sensitivity to
$\Delta{m}^2_{41}$ worsens decreasing
$\sin^2 2\vartheta_{ee}$.
Indeed,
for small values of
$\sin^2 2\vartheta_{ee}$
we have
$|U_{e4}|^2 \simeq \sin^2 2\vartheta_{ee} / 4$
and the contribution of $m_{4}^2 \simeq \Delta{m}^2_{41}$
to the spectrum (\ref{lambdasterile2}) is suppressed.
On the other hand,
the sensitivity to $m_{4}^2 \simeq \Delta{m}^2_{41}$
for $\sin^2 2\vartheta_{ee} = 1$
is only slightly worse of that for $m_{\nu}^2$ in the three-neutrino
mixing case discussed in Section~\ref{sec:3nu},
because $\sin^2 2\vartheta_{ee} = 1$ corresponds to
$|U_{e4}|^2 = 1/2$.

In Fig.~\ref{fig:sensitivitysterile} we also depicted
the region allowed at
95.45\% C.L.
by a global fit of short-baseline neutrino oscillation data \cite{Giunti:2013aea,Gariazzo:2015rra}
and the
95.45\% C.L.
allowed regions
obtained by restricting the analysis to the data of $\nu_{e}$ and $\bar\nu_{e}$ disappearance experiments
\cite{Giunti:2012tn,Giunti:2012bc},
taking into account the
Mainz
\cite{Kraus:2012he}
and
Troitsk
\cite{Belesev:2012hx,Belesev:2013cba}
bounds.
These last regions are interesting because it is possible that the
disappearance of $\nu_{e}$ and $\bar\nu_{e}$
indicated by the reactor and Gallium anomalies will be confirmed by
the future experiments
whereas the LSND anomaly will not.

From Fig.~\ref{fig:sensitivitysterile} one can see that the
$\nu_{e}$ and $\bar\nu_{e}$ disappearance region is wider than the
globally allowed region and extends to values of
$\Delta{m}^2_{41}$
as large as about 80 eV$^2$.
Hence, it can be partially explored by the ECHo-1M experiment,
which is expected to have a statistics of
$N_{\text{ev}} \simeq 10^{14}$.

Figure~\ref{fig:sensitivitysterile} shows that
in order to explore
the region which is allowed by the global fit of short-baseline neutrino oscillation data
it will be necessary to make a $^{163}$Ho experiment with a statistics
$N_{\text{ev}} \gtrsim 10^{16}$.
One can also see that
an $^{163}$Ho experiment with this statistics will be competitive with the
KATRIN experiment
\cite{Mertens-TAUP2015},
a result that is consistent with that for the sensitivity on $m_{\nu}$
in the standard framework of three-neutrino mixing discussed at the end of
Section~\ref{sec:3nu}.

Figure~\ref{fig:sensitivitysterile} also shows that
the exploration of the
small-$\Delta{m}^2_{41}$
regions allowed by the $\nu_{e}$ and $\bar\nu_{e}$ disappearance data
will require a statistics as high as
$N_{\text{ev}} \approx 10^{18}$.

\section{Conclusions}
\label{sec:conclusions}

In this paper we presented the results of
an analysis of the sensitivity of $^{163}$Ho experiments
to neutrino masses
considering first the effective neutrino mass $m_{\nu}$
in the standard framework of three-neutrino mixing
(see Eq.~(\ref{mnu}))
and then
an additional mass $m_{4}$ at the eV scale
in the framework of 3+1 neutrino mixing
with a sterile neutrino.
We considered the experimental setups
corresponding to the two
planned stages of the ECHo project,
ECHo-1k and ECHo-1M \cite{Gastaldo:2013wha,Hassel:2016echo}.

We found that
the ECHo-1k experiment
can reach a sensitivity to $m_{\nu}$
of about 6.5 eV at $2\sigma$
with a total statistics of $N_{\text{ev}} \simeq 10^{10}$,
an energy resolution $\Delta{E}_{\text{FWHM}} \simeq 5$ eV
and
a pileup fraction
$f_{\text{pp}} \simeq 10^{-6}$.
Although this sensitivity
is still not competitive with that of tritium-decay experiments,
it will represent an improvement of
more than one order of magnitude with respect to
the current limit
$m_{\nu} < 225$ eV at $2\sigma$ \cite{Springer:1987zz}
obtained with a
$^{163}$Ho electron capture experiment.
We also found that
the ECHo-1k experiment will not allow to put more stringent limits on
the mass and mixing of $\nu_{4}$
than those already obtained in the
Mainz
\cite{Kraus:2012he}
and
Troitsk
\cite{Belesev:2012hx,Belesev:2013cba}
experiments.

According to our estimation, the second stage of the ECHo project,
ECHo-1M,
can reach a sensitivity to $m_{\nu}$ of about 0.7 eV at $2\sigma$
with
$N_{\text{ev}} \simeq 10^{14}$,
$\Delta{E}_{\text{FWHM}} \simeq 2$ eV
and
$f_{\text{pp}} \simeq 10^{-6}$.
This result will narrow the gap between
the sensitivities of
tritium-decay experiments and $^{163}$Ho electron capture experiments.
Indeed,
0.7 eV is smaller than the current upper limit of about 2 eV at $2\sigma$
obtained in the
Mainz \cite{Kraus:2004zw} and Troitsk \cite{Aseev:2011dq} experiments
and it is not too far from the expected sensitivity of about 0.2 eV
of the KATRIN experiment
\cite{Osipowicz:2001sq,Mertens-TAUP2015}.

We found that the ECHo-1M
experiment will be sensitive
to the large-$\sin^2 2\vartheta_{ee}$ and large-$\Delta{m}^2_{41}$
part of the region in the
$\sin^2 2\vartheta_{ee}$--$\Delta{m}^2_{41}$
plane
which is allowed by the data of short-baseline
$\nu_{e}$ and $\bar\nu_{e}$ disappearance experiments
\cite{Giunti:2012tn,Giunti:2012bc},
taking into account the
Mainz
\cite{Kraus:2012he}
and
Troitsk
\cite{Belesev:2012hx,Belesev:2013cba}
bounds.
However, it cannot explore
the region allowed by the global fit of short-baseline neutrino oscillation data \cite{Giunti:2013aea,Gariazzo:2015rra}.

According to our calculations,
a $^{163}$Ho electron capture experiment
with
$\Delta{E}_{\text{FWHM}} \simeq 2$ eV
and
$f_{\text{pp}} \simeq 10^{-6}$
will be competitive with the KATRIN tritium-decay experiment
\cite{Osipowicz:2001sq,Mertens-TAUP2015}
by reaching a statistics of $N_{\text{ev}} \approx 10^{16}$.
Such an experiment
will cover a large part of the region in the
$\sin^2 2\vartheta_{ee}$--$\Delta{m}^2_{41}$
plane
which is allowed by the data of short-baseline
$\nu_{e}$ and $\bar\nu_{e}$ disappearance experiments
and
the
large-$\sin^2 2\vartheta_{ee}$ and large-$\Delta{m}^2_{41}$
part of
the region allowed by the global fit of short-baseline neutrino oscillation data.

In order to explore all the region allowed by the global fit of short-baseline neutrino oscillation
it will be necessary to have
a statistics of $N_{\text{ev}} \approx 10^{17}$
and
to cover all the region allowed by the data of short-baseline
$\nu_{e}$ and $\bar\nu_{e}$ disappearance experiments
a statistics of $N_{\text{ev}} \approx 10^{18}$
will be needed.
These large event numbers seem unreachable now,
but we think that we should be optimistic,
taking into account that the development of $^{163}$Ho electron capture experiment
is only at the beginning.

\begin{acknowledgments}
We would like to thank
A. De Rujula,
A. Faessler,
M. Lusignoli,
A. Nucciotti,
T. Schwetz
for fruitful discussions.
L.G. acknowledges the
support by the DFG Research Unit FOR 2202 ``Neutrino Mass Determination by Electron Capture in 163Ho, ECHo'' (funding under GA 2219/2-1).
The work of C.G.
was partially supported by the research grant {\sl Theoretical Astroparticle Physics} number 2012CPPYP7 under the program PRIN 2012 funded by the Ministero dell'Istruzione, Universit\`a e della Ricerca (MIUR).
E.Z. thanks the support of funding grants 2013/02518-7 and 2014/23980-3, S\~ao Paulo Research Foundation (FAPESP).
\end{acknowledgments}


\end{document}